\documentclass[sigconf]{acmart}
\usepackage{natbib}
\usepackage{enumitem}
\usepackage{tcolorbox}
\usepackage{mdframed}
\usepackage{booktabs}

\AtBeginDocument{%
  \providecommand\BibTeX{{%
    \normalfont B\kern-0.5em{\scshape i\kern-0.25em b}\kern-0.8em\TeX}}}

\copyrightyear{2020}
\acmYear{2020}
\setcopyright{acmcopyright}\acmConference[ESEM '20]{ACM / IEEE International Symposium on Empirical Software Engineering and Measurement (ESEM)}{October 8--9, 2020}{Bari, Italy}
\acmBooktitle{ACM / IEEE International Symposium on Empirical Software Engineering and Measurement (ESEM) (ESEM '20), October 8--9, 2020, Bari, Italy}
\acmPrice{15.00}
\acmDOI{10.1145/3382494.3422172}
\acmISBN{978-1-4503-7580-1/20/10}

\begin{document}

\title{Profiling Developers Through the Lens of Technical Debt}

\author{Zadia Codabux}
\authornotemark[1]
\affiliation{%
  \institution{University of Saskatchewan}
  }
\email{zadiacodabux@ieee.org}

{
\author{Christopher Dutchyn}
\affiliation{%
 \institution{University of Saskatchewan}
  }
\email{dutchyn@cs.usask.ca}
}

\renewcommand{\shortauthors}{Codabux et al.}

\begin{abstract}
\textbf{Context:} Technical Debt needs to be managed to avoid disastrous consequences, and investigating developers' habits concerning technical debt management is invaluable information in software development.
\textbf{Objective:} This study aims to characterize how developers manage technical debt based on the code smells they induce and the refactorings they apply. 
\textbf{Method:} We mined a publicly-available Technical Debt dataset for Git commit information, code smells, coding violations, and refactoring activities for each developer of a selected project. 
\textbf{Results:} By combining this information, we profile developers to recognize prolific coders, highlight activities that discriminate among developer roles (reviewer, lead, architect), and estimate coding maturity and technical debt tolerance.
\end{abstract}




\maketitle

\section{Introduction}
\label{sec:intro}

Open Source Software (OSS) platforms have thrived over the last decade. The latest "State of the Octoverse" report\footnote{https://octoverse.github.com/} by GitHub\footnote{https://github.com/}, one of the world's most popular and leading software development platforms for both open source and private project repositories, reports that 44+ million projects were developed by 40+ million developers from all over the world. OSS development is typically a collaborative process. Most OSS developers are skilled and experienced professionals with an average of 11.8 years of software development experience. About 50\% of the developers have acquired formal training in computer science or related domain, and the rest are self-taught or learned the skills on the job~\cite{lakhani2003hackers}. 

Different developers have different weaknesses and tolerances for Technical Debt (TD). TD is a metaphor introduced by Ward Cunningham~\cite{cunningham1992wycash} to help reason about the long-term costs incurred by taking shortcuts in software development. However, when TD is not managed, it can impede software development in multiple ways ranging from quality issues to loss of a system. One of the indicators of TD is \textit{code smells}~\cite{zazworka2014comparing}, a term coined by Kent Beck, which refer to surface indication of potential design weaknesses in source code~\cite{fowler2018refactoring}. One of the ways to manage TD is refactoring~\cite{codabux2014quality}. Refactoring is a technique for restructuring code by improving its internal structure without affecting its external behavior~\cite{fowler2018refactoring}. Therefore, establishing the developers' shortcomings with respect to TD can help focus scrutiny on portions of source code that may be more likely to contain flaws or debt, both by the coder during software development and by reviewers when they check the code during code reviews. 

This study aims to profile developers with respect to technical debt by mining developers' commit actions, their code smells, and refactoring activities. This could help in team selection and organization, task assignments such as complex refactoring or code reviews, and pair programming sessions. We mined the publicly available TD dataset~\cite{lenarduzzi2019technical} to obtain the commit information from GitHub repositories, code smells, and coding violations from SonarQube\footnote{https://www.sonarqube.org/} and Ptidej~\cite{gueheneuc2007ptidej}, and refactoring information from Refactoring Miner~\cite{tsantalis2018accurate}. We elaborate on the TD dataset and tools in Section~\ref{sec:background}.

The contributions of this study are: 
\begin{enumerate}[noitemsep,topsep=0pt]
        \item We profile developers with respect to technical debt based on the code smells they induce and close and the type of refactoring they perform.
        \item An updated TD dataset (in the .db format) with views for extracting the raw data for profiling developers, constructed using SQL queries.
        \item The mapping of SonarQube coding violations to the most common code smells by Fowler~\cite{fowler2018refactoring} and the translation of these rules violations to bad practices, in the format of a CSV file (codified.csv in the replication package (Section~\ref{sec:rep_pack})).
\end{enumerate}

The rest of the paper is organized as follows. Section~\ref{sec:background} describes the terminologies associated with the tools in the TD dataset used throughout this study. Section~\ref{sec:related_work} highlights the relevant related work about developers' characterization. Section~\ref{sec:methodology} focuses on the methodology of the study, including the research goal, questions, study design, and the data analysis process. Section~\ref{sec:results} provides insights into the results. Section~\ref{sec:discussion} presents the discussion. Section~\ref{sec:threats} lists threats to the validity of this study. Section~\ref{sec:conclusion} concludes and outlines future work.

\section{Background}
\label{sec:background}

In this section, we elaborate on the tools and associated terminologies that we use in this paper. 

Git distinguishes between these different developers' roles\footnote{https://bit.ly/3ezc5GH}:
\begin{itemize}
    \item \emph{Author} is the person who originally wrote the work
    \item \emph{Committer} is the person who last applied the work
\end{itemize}

Refactoring Miner is an open-source tool that mines refactoring operations from the git commits of Java projects. Some examples of refactoring types extracted are: \emph{Extract Interface}, \emph{Move Method}, and \emph{Extract Method}~\cite{fowler2018refactoring}.

SonarQube is a web-based open-source platform to analyze source code to detect violations to its pre-defined rules. SonarQube defines three types of issues\footnote{https://bit.ly/2Cjp1Dz}:
\begin{itemize}
\item Bug: an error that breaks the code; fix immediately
\item Security Vulnerability: a point in code that's open to attack
\item Code Smell: a maintainability issue that makes the code confusing and difficult to maintain
\end{itemize}

SonarQube has more than 500 rules\footnote{https://rules.sonarsource.com/java} for Java issues relating to different aspects of source code. SonarQube also classifies the rules into five severity levels: Blocker, Critical, Major, Minor, and Info. In this study, we only considered the SonarQube rules violations (which SonarQube refers to as "code smells"), but we mapped the rules violations to Fowler's widely-known code-smell classification~\cite{fowler2018refactoring} and bad coding practices. Henceforth, when we mention code smells in this study, we are referring to Fowler's classification. Many bad practices are not categorized in the SonarQube data. In fact, when delving into the TD dataset, we discover that more than 50\% of the SonarQube rules violations end up uncategorized. 

\section{Related Work}
\label{sec:related_work}
Here, we explore literature for studies on profiling OSS developers.

Yang et al.~\cite{yang2020developer} proposed a developer portrait model based on three characteristics: personal information, programming skills, and contributions from the Open Source Platforms GitHub, GitLab, Gitee, and Bitbucket. Then, they successfully applied the model to code recommendation and programming tasks assignments. 
Avgustinov et al.~\cite{avgustinov2015tracking} analyzed the static analysis violations over the revision histories of the open-source projects: Hadoop Common, MongoDB, Spark, and Gaia. They tracked down which developers were responsible for introducing and fixing these violations to build developers' profiles about their coding habits. 
Leveraging the information on how defect is prioritized by software developers in defect repositories for the Open Source Systems Eclipse and Mozilla is the essence of the study conducted by Xuan et al.~\cite{xuan2012developer}. The authors modeled the developer's bug prioritization to assist in tasks such as bug triage, severity identification, and reopened bug prediction. 
Along the same lines, Xia et al.~\cite{xia2013accurate} analyzed bug reports from GCC, OpenOffice, Mozilla, Netbeans, and Eclipse to recommend developers for bug fixing. 
Amor et al.~\cite{amor2006effort} proposed to use information from source code management systems, mailing lists archives, and bug tracking systems to characterize developer activities for more accurate effort and cost estimations in OSS. 
A study by Hauff et al.~\cite{hauff2015matching} extracted GitHub user-profiles, specifically, their README.md files, to match developers' information with descriptions from job advertisements. 
Li~\cite{li2019research} conducted a study focused on the analysis of developers' behaviors related to technical debts. Using clustering on Git commits, faults and SonarQube issues, including vulnerabilities, the author came up with the following categories of developers: Veterans, Vulnerability Creators, and Fault Inducers/
Commoners.

Although there are a few studies that explore the characteristics of developers in OSS based on artifacts (e.g., as defect information and source code) for various purposes such recommending specific tasks (e.g., bug fixing) or for mapping developer profiles to jobs, to the best of our knowledge, profiling developers from a technical debt perspective is yet to be explored from a combined code smell and refactoring perspective. 

\section{Methodology}
\label{sec:methodology}

\subsection{Goal}
\label{sec:goal}

The goal of this study is to mine developers' actions from code smells and refactoring information to profile developers from a technical debt perspective. In so doing, we will identify the developers who are superior at managing tasks, fixing bugs, and refactoring code. This developer characterization will help inform decisions regarding team selection and organization, task assignments, and pair programming sessions. We propose the following Research Question (RQ) to gather information to profile developers. \\
\textbf{
What do the developers' activities regarding inducing and refactoring code smells and bad practices tell us about their technical debt management?
}
\subsection{Study Design}
\label{sec:design}

The data used in this study is from the publicly available Technical Debt Dataset (TD dataset), downloaded as a database, which consists of 33 Java projects from the Apache Software Foundation\footnote{https://www.apache.org/} repository. The TD dataset contains commit information from the GitHub commit logs, refactoring extracted using Refactoring Miner, code smells, and programming violations from SonarQube and Ptidej. The projects in the TD dataset are more than three years old, contained at least 500+ commits and 100+ classes. \emph{Because the dataset is public domain and contains full developer identities, our anonymization can be trivially reversed.}  

\begin{table}[!htb]
\begin{minipage}{\linewidth}
\centering
\resizebox{\linewidth}{!}{
\begin{tabular}{ll}
\toprule
\textbf{Table Name} & \textbf{Description}  \\
\midrule
Git Commits & Commit information retrieved from the commit log \\
Sonar Measures & SonarQube measures for the commits \\
Sonar Issues & SonarQube issues, antipatterns and code smells\\
Refactoring Miner & List of refactoring activities\\
\bottomrule
\end{tabular}
}
\end{minipage}
\vspace{-2ex}
\caption{TD dataset SQL Tables\vspace{-5ex}}
\label{tbl:tdd_tables}
\end{table}

For this study, we selected the project Apache Beam (\texttt{beam})\footnote{https://beam.apache.org/} which is one of the largest projects in the TD dataset. \texttt{beam} is a unified programming model used to implement batch and streaming data processing pipelines. The data for \texttt{beam} has been collected over 5 years (2014-2019) and consist of 22,300+ commits, 10,100+ refactoring, 74,400+ technical debt items. Since the focus of this study is on profiling developers from a refactoring and code smells perspective, we only used the related tables described in Table~\ref{tbl:tdd_tables}.

\subsection{Data Analysis}
\label{sec:data_analysis}
Beginning with the TD dataset, we construct SQL queries to illuminate our research question, with summary statistics presented graphically. The data slices into three different perspectives:
\begin{enumerate}
    \item The \textbf{commits} (\texttt{GIT{\textunderscore}COMMITS}) will allow us to characterize the distribution of authorship, identify prolific code authors, and recognize central authorities who commit changes.
    \item Details of statically-recognized programming concerns using SonarQube rules, which we categorized as \textbf{code smells} and \textbf{bad practices} along with the author of the defective code, the author and the committer of the repaired code, and the status of the concern.  
    \item \textbf{Refactoring} data allows us to trace standard refactorings ~\cite{fowler2018refactoring} back to the original author; ranging from simple \emph{RenameVariable} tasks to complex \emph{ExtractSuperclass} actions. 
\end{enumerate}

\noindent\textbf{Commits\ \ }
\label{sec:commits}
This data, displayed in Figure~\ref{fig:handles}, tells us about authorship and commits for each developer, and we recognize the long-tail distribution and anomalies, which help us recognize specific developers. 
Also, the data only contains commits to the \texttt{main} branch, so we do not see prototyping or alternate exploration activities. However, the commits are tagged by \textit{author}, so we know who constructed the code. The developer is frequently different from the \textit{committer}, a point for discussion later. 
 
There are several obstacles which introduce noise and complexity to the data supporting the analyses:
\begin{itemize}
    \item Some individuals use more than one distinct but unambiguous \emph{name} in the database---anonymized e.g.:~\texttt{jsmith} and \texttt{Jon Smith} and \texttt{Jonathan Smith}
    \item Some entries are unambigous email addresses or user-ids---anonymized e.g.:~\texttt{nasa} and \texttt{nasa@nasa.gov}
    \item Other short forms, are ambiguous and cannot be resolved to an individual in the project---anonymized e.g.:~\texttt{jsmith} might be \texttt{John Smith}, \texttt{Jon Smith}, or \texttt{Jane Smith}.
    \item Some commits are not tagged or use a non-descript e.g. \texttt{GitHub} or \texttt{MergeBot}
\end{itemize}

We enumerate these various names into the table \texttt{BEAM{\textunderscore}HANDLES}, which contains rows, each with an \texttt{asEntered} name-string, and a \texttt{handle} for each; the handles are then anonymized into strings such as \texttt{Dev1}.  Each of these fields are marked as case-indifferent\footnote{This is the SQL-Lite term for case-insensitive.}, and so collating sequences will put \texttt{Bill} near \texttt{bill} and \texttt{BILL}. By joining the \texttt{HANDLE} table on the data-source's name-string, and extracting the \texttt{handle}, we can gather the various different names for a single developer to one canonical \texttt{handle}.  This is seen in our queries in the extended SQL-Lite database as:
\begin{verbatim}
    select ..., h.handle, ... from ...
       join BEAM_HANDLES h on h.asEntered = ...
\end{verbatim}

In addition, our \texttt{BEAM{\textunderscore}HANDLES} includes counts for the number of \texttt{GIT{\textunderscore}COMMITS} with the \texttt{asEntered} name-string as the \emph{author} and \emph{committer}, as well as counts of SonarQube rule violations closed.

For the remaining three data-sources, just as above, we construct similar queries to select, combine, and summarize the different elements; then create informative charts to illustrate our learnings.

\noindent\textbf{Code Smells} and \textbf{Bad Practices\ \ } 
In the TD dataset, \texttt{SONAR{\textunderscore}MEASURES} (almost) joins on the \texttt{commitHash} with \texttt{GIT{\textunderscore}COMMITS}, connecting SonarQube data with commits. Unfortunately, the \texttt{SONAR} tables  truncate the hash by one leading character, but entries still remain in a 1-1 correspondence: every row in \texttt{MEASURES} has a unique \texttt{COMMIT} with a longer hash.  Therefore, we extend \texttt{{\textunderscore}BEAM{\textunderscore}GIT{\textunderscore}COMMITS} with an additional computed field, \texttt{shortHash}, that contains the truncated \texttt{commitHash}, which we use to build 
\texttt{BEAM{\textunderscore}SONAR{\textunderscore}MEASURES{\textunderscore}COMMITS}. Then, we can join that view with the\texttt{{\textunderscore}BEAM{\textunderscore}SONAR{\textunderscore}ISSUES} to connect an issue \texttt{{\textunderscore}DETAIL} to each closing commit and thus to each author and committer for that issue.

\begin{figure}
    \centering
    \includegraphics[width=0.5\textwidth]{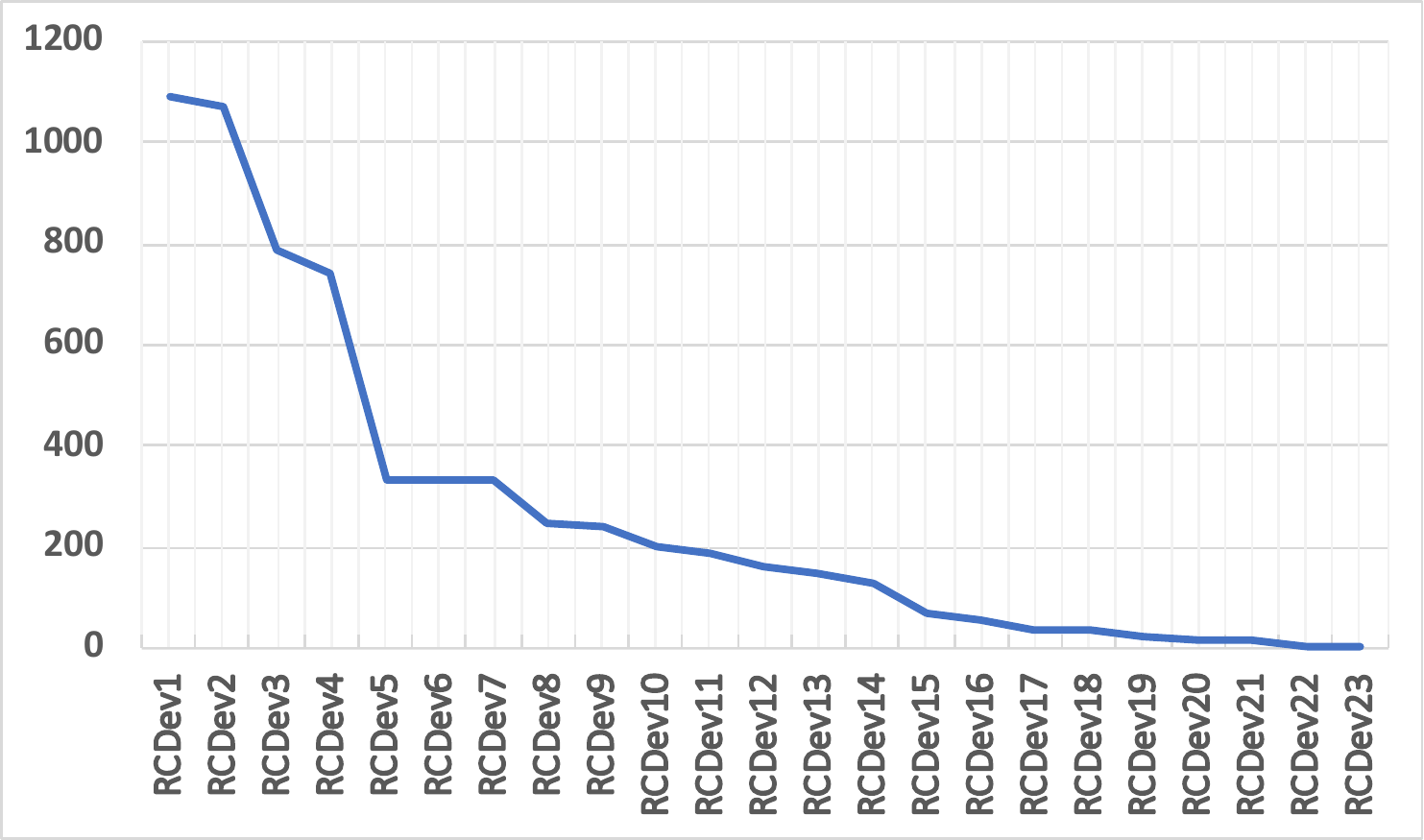}
    \vspace{-5ex}
    \caption{Count of Refactoring Commits\\[-2ex]}
    \label{fig:refcnt}
\end{figure}

\noindent\textbf{Refactoring Activities\ \ }
Many of the Git commits have refactoring descriptions attached, using the \texttt{commitHash} as the join field.  This allows us to compute summary statistics for the frequency of the various kinds of refactoring as a cross-table by author in Figure~\ref{fig:refcnt}. But first, we recognize that the most prolific refactorers are also the authors tending to the right-hand side (expert) of the SonarQube data in that figure.

\subsection{Replication Package}
\label{sec:rep_pack}
The full set of queries and views along with the data underlying our study is publicly available\footnote{https://bit.ly/3h7PGSb}. Specifically, we provide the SQL-Lite database with executable views for all tables, various spreadsheets formatting the diagrams, and the codified code smells file.

\section{Results}
\label{sec:results}

This section reports findings for our research question based on the three perspectives in Section~\ref{sec:data_analysis}. 

\begin{figure}
    \centering
    \includegraphics[width=0.5\textwidth]{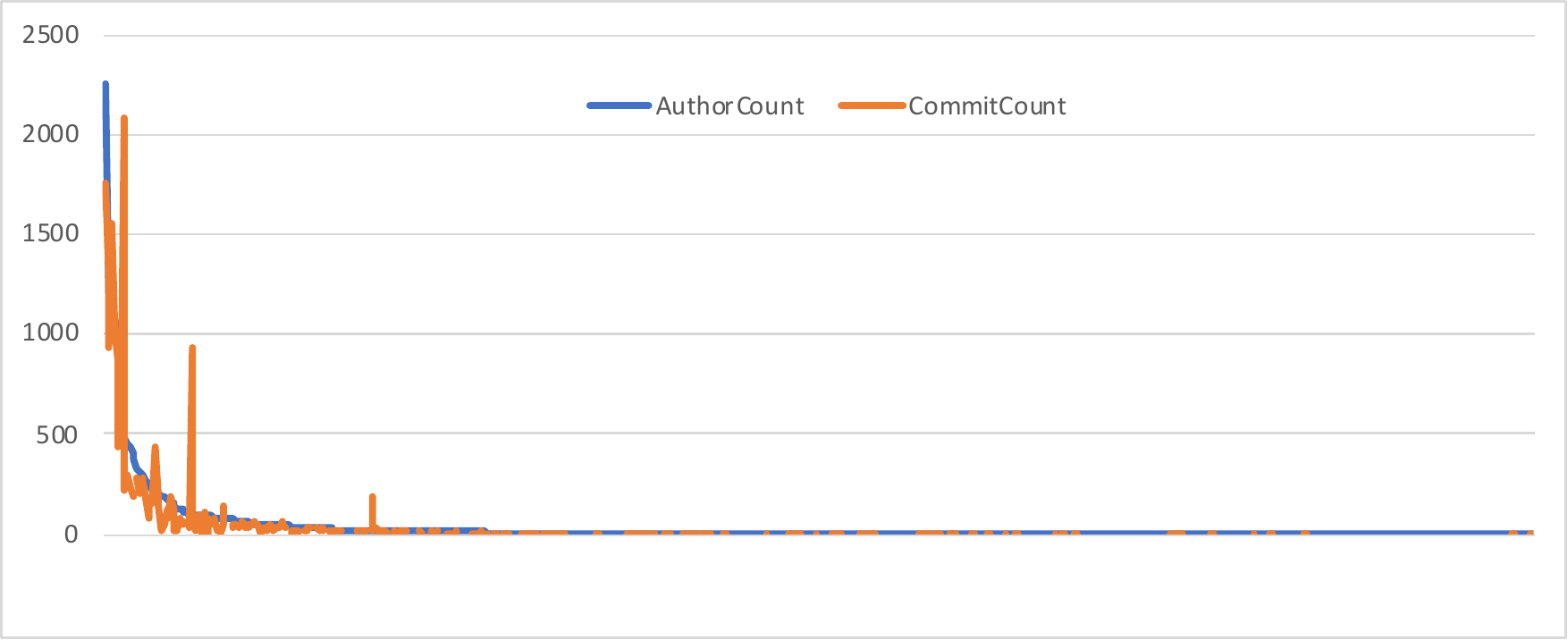}
    \vspace{-5ex}
    \caption{Distribution of Authoring and Committing\\[-2ex]}
    \label{fig:handles}
\end{figure}

\noindent\textbf{Commits\ \ }
From the commit view in Section~\ref{sec:commits}, we identify 754 unique name-strings, and this reduces down to 643 unique handles, shown in our \texttt{BEAM{\textunderscore}HANDLES{\textunderscore}SUMMARY} view.  By sorting the view by descending author count, Figure~\ref{fig:handles} displays the long-tail distribution of authors as a blue-line from the highest at 2256 on the left, to the lowest at 1 on the right.  Overlaying the committer count for each handle, as an orange-line superimposed in the same figure, we see that commits tend to follow the same curve, but with a few notable differences, which highlight specific developers.

Most committers are authors, and over-committers stand out as blips where the orange (CommitCount) line jumps up above (or below) the blue-line (AuthorCount).  Four authors stand out:
\begin{itemize}
    \item at rank 9 with 506 authorships, 2085 commits,
    \item at rank 14 with 377 authorships, but no commits,
    \item at rank 39 with 98 authorships, and many (929) commits, and
    \item at rank 119 with 19 authorships, and many (188) commits.
\end{itemize}

The small blip at rank 23 in Figure~\ref{fig:handles} is \texttt{MergeBot} with 218 authorships and 429 commits.  This is a generic handle for some automated process and does not inform us further. Overall, the top 22 handles represent 63\% (14,106 of the 22332) of all authorships; and 70\% (13,060 of 18,550) of all commits to the main branch.  Therefore, we will tend to focus on the top 22 handles, and the two other over-committers (at ranks 39 and 119); and these are noted in the \texttt{BEAM{\textunderscore}HANDLES} table as having an entry in a column, \texttt{significantContributor} which \texttt{IS NOT NULL}.

\noindent\textbf{Code Smells and Bad Practices\ \ }
Summarizing our statistics for SonarQube rule violations (which include Fowler code-smells and other bad practices), we find 78,642 instances of an author making a change that closes a recognized SonarQube rule violation, with 2602 distinct messages.  We extracted all 2602  unique \texttt{Detail} messages from this table, and manually marked them with with a \texttt{Type} and \texttt{SubType}, as described in \texttt{codified.csv} in our replication package.  Our summary view, \texttt{BEAM{\textunderscore}SONAR{\textunderscore}ISSUES{\textunderscore}SUMMARY}  shows the details; briefly, there are three \texttt{Types}:
\begin{itemize}
    \item Anti-Pattern (with one SubType, \emph{Anti-Singleton})
    \item Bad Practice (with numerous SubTypes, including our own)
    \item Code Smell (with 12 standard SubTypes).
\end{itemize}

We summarized SonarQube messages against significant contributors by handle, as our \texttt{BEAM{\textunderscore}SONAR{\textunderscore}MESSAGES{\textunderscore}BY{\textunderscore}AUTHOR} view.

\begin{figure}
    \centering
    \includegraphics[width=0.5\textwidth]{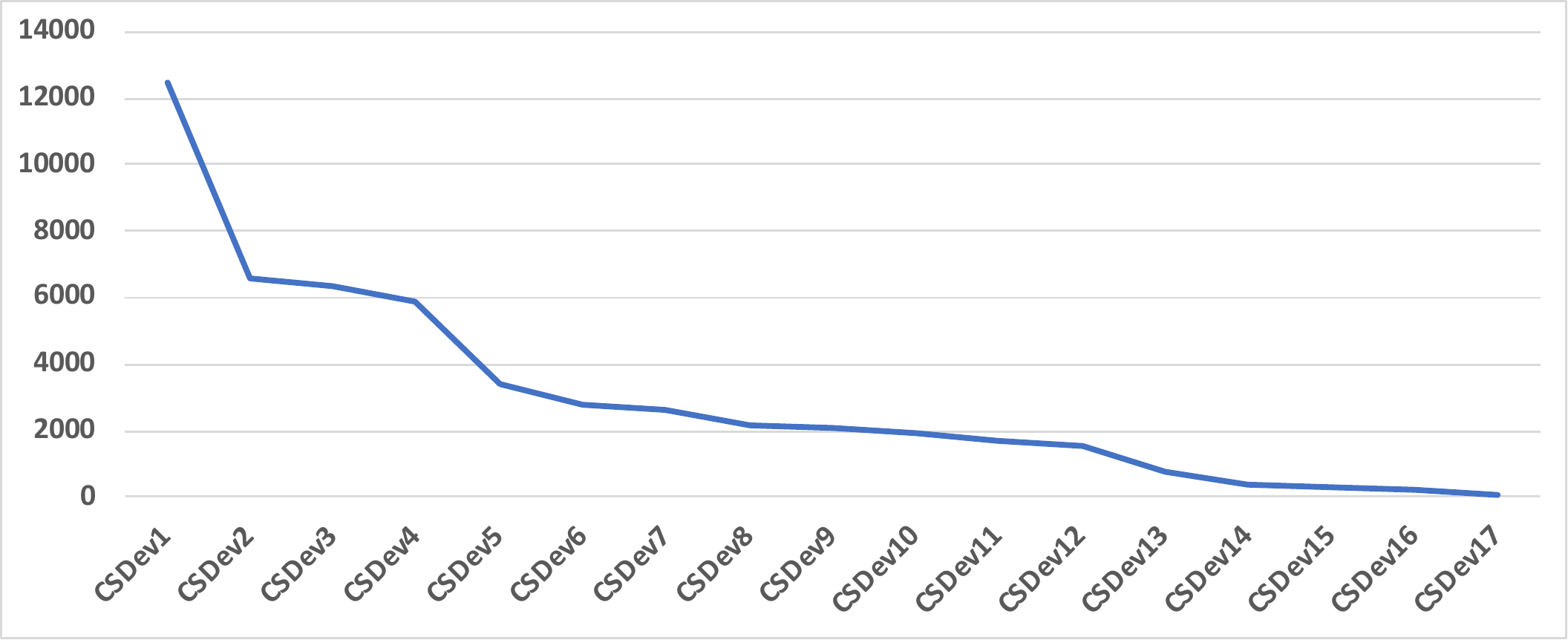}
    \vspace{-5ex}
    \caption{Count of Sonar Violations Resolved by Author\\[-2ex]}
    \label{fig:sonarcnt}
\end{figure}

Seventeen of our significant contributors authored changes that closed SonarQube rule violations.  The most prolific contributor removed 12,500 SonarQube violations of all types. 
 All told, seventeen authors repaired 51,187 instances.  The \texttt{BEAM{\textunderscore}SONAR{\textunderscore}MESSAGES{\textunderscore}\\SUMMARY} computes these values; and Figure~\ref{fig:sonarcnt} displays the curve.

\begin{figure*}
    \includegraphics[width=\textwidth,trim=0 200 0 172,clip]{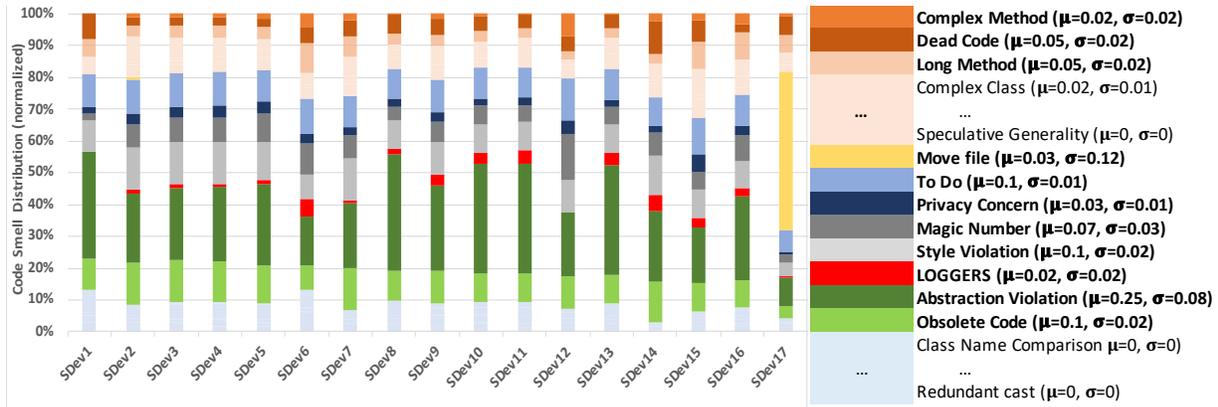}
    \vspace{-5ex}
    \caption{Sonar Violations Distribution by Author and Type\\[-2ex]}
    \label{fig:cross}
\end{figure*}

We extracted those records to a \texttt{csv} file for import into a presentation tool. Using simple text-processing tools, we converted the row representation into a cross-table in graphical format, c.f.~ Figure~\ref{fig:cross}.  The graph has rows and columns reordered to emphasize interesting details, deferred to the Discussion section.
In this figure, the total number of commits is normalized to 100\%, and the variation among the types and subtypes is emphasized.  Note that the order of authors is anonymized separately and in a different order from previous figures.  Each vertical bar corresponds to a single author (captioned along the bottom) and is divided into distinguishing color bands based on type and subtype.  The order of the bands is identical across all authors.

The sole \textit{AntiPattern} type (with unique \textit{Anti-Singleton} subtype) is so small as to be indistinguishable in black at the top of each bar: indeed, there were only 17 instances of this among 51,187 SonarQube violations. For the remaining SonarQube rule violations, we stacked code smells above bad practices. Twelve of the recognized code smells~\cite{fowler2018refactoring} were identified in the data, but only three showed substantial difference across the developers: \textit{LongMethod}, \textit{DeadCode}, and \textit{ComplexMethod}.
Therefore, we color-coded the three interestingly-varied code-smells in different bright-orange and merged the other nine into dim-orange.

\begin{figure*}
    \includegraphics[width=\textwidth,trim=0 190 0 192,clip]{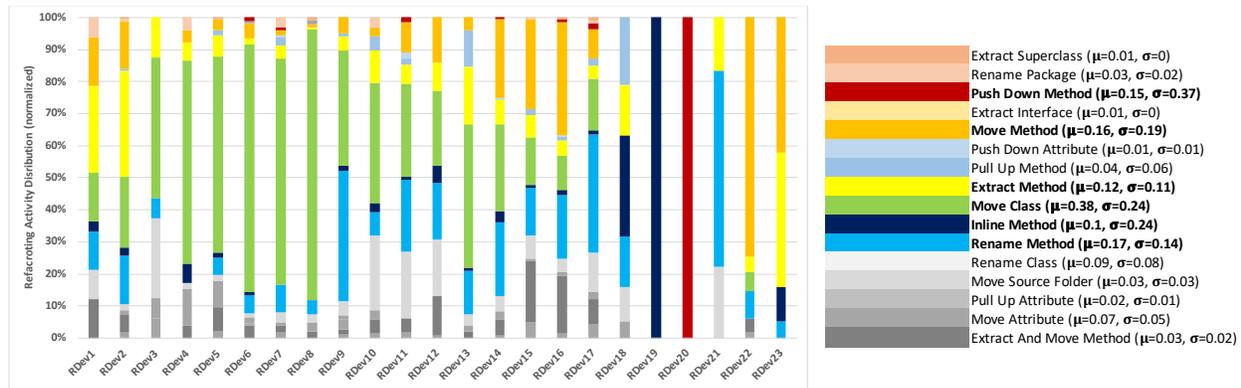}
    \vspace{-5ex}
    \caption{Refactorings by Type and Author\\[-2ex]}
    \label{fig:refactor}
\end{figure*}

In the same way, we examined the bad practices data, and there were 34 distinct sub-types; but only eight showed up more than $1\%$ of the time, and all of those had noticeable variation (as measured by standard deviation); they are bolded in Figure~\ref{fig:refactor}.

The other 26 subtypes of bad practices were present in $1\%$ of the violations, and standard deviation of at most $1\%$, and merged together into the very dim blue at the bottom.  We note a few of these:
\begin{itemize}
    \item Unlike most practices, only three developers engaged in \textit{MoveFile} (the yellow band), and one of them performed almost all of the file moves, and almost half of his work was resolving those rule violations.  
    \item Abstraction violations were almost one-quarter of all rule violations but tended to dominate the others who also had higher levels of style violations, obsolete code fixes, and magic numbers. 
\end{itemize}
We defer the interpretation of these two points to a later section.

\noindent\textbf{Refactoring Activities\ \ }
With the cross-table query, we get 23 distinct significant contributors and a categorization of their commits into 16 distinct types.  Some of them are small contributions (less than 10\% of the instances, and single-digit variances), but six have some presence, and two more are close.  Figure~\ref{fig:refactor} shows the distribution of activity, much as we did previously.

Of note in the table,
\begin{itemize}
    \item Two authors performed one refactoring exclusively: \emph{Inline Method} and \emph{Push Down Method} respectively
    \item And except for those two, everyone else employed \emph{Extract Method}, a relatively sophisticated refactoring type
    \item \emph{Move Class} was a common refactoring for most other authors, and in some cases, a substantial majority of their actions
    \item Only five authors employed \emph{Move Method}, but they used it extensively ($25\%$ or more).
\end{itemize}

\section{Discussion}
\label{sec:discussion}
Our results offer numerous opportunities for discussion.  We have confidence that our analysis offers a fair assessment of the available data, regardless of our choice to focus on a small subset of contributors.  Overall, the top 22 handles represent 63\% (14,106 of the 22,332) of all authorships; and 70\% (13,060 of 18,550) of all commits to the main branch.  This gives us a good understanding of the various developers involved in the \texttt{beam} project.

\noindent\textbf{Profiling Developers\ \ \ }
We do not produce a profile document for each developer; instead, we perform the act of profiling, recognizing the similarities and differences among developers and relating those to a level of coding maturity and coding specialization.  We profile developers by author and commit activity, recognizing those with unusual commit counts compared to authoring---these individuals appear to be distinguishable code-reviewers, technical leads, or project managers.  It also allows us to discriminate different kinds of code authors: some perform global changes such as moving code among packages---these might be architects.  Others create or close large numbers of \textit{ToDo} items---perhaps quality assurance personnel.  One focused exclusively on pushing methods down: dealing with \emph{Generalization} problems.  He would need to carefully consider the risks of running general code (a \texttt{super} method) over methods specialized to different classes---i.e., a developer with in-depth knowledge of the package.  Another simply inlined methods: a method-composition act yielding performance optimization.  This is a task for a developer who is confident that the original method is superfluous: one who understands when a method abstraction is unnecessary and will not add excessive complexity---i.e., someone with experience, knowledge of the code, and good intuitions about code complexity. Some others stand out for large numbers of Abstraction Violations, Complex code, obsolete code (perhaps cribbed from StackOverflow), or trivial Style-Guide or Magic-Number violations---often a hallmark of immature developers.  Developers with less than 80\% of their SonarQube rule violations identified as bad practices tended to introduce more code smells (except for the file-mover)---developers with fewer bad practices, especially of the more trivial sort, seem to tolerate larger amounts of code smell.  This profiling exercise suggests that mature developers produce fewer bad practices, but admit larger amounts of technical debt.

\section{Threats to Validity}
\label{sec:threats}
This is a preliminary study based on one system, \texttt{beam}, specific to the Apache foundation and written in Java. Although it is a large project over multiple years, we cannot generalize our results to other projects, both OSS and industrial. However, this study serves as a basis to expand this research to other systems. The TD dataset provides snapshots of projects over a specific timeframe, and there might be some missing data. However, the data for \texttt{beam} spans since 2014, which coincides with the first release of an open SDK implementation by Google. So we can assume that we have a fairly complete picture of the project over the TD dataset timeline. 

We used the data from the TD dataset database, which uses specific tools and algorithms. Most of the tools were used with the standard configuration. So we might have missed some of the data that the tools might not have reported. Most of the data in the Git commits are self-reported by the developers and might not be an accurate reflection of the actual data due to a lack of knowledge and experience to identify the correct components or assign the right labels. 

The mapping of the SonarQube rule violations to the more commonly known code smells by Fowler, and bad practices (captured in our \texttt{codified.csv}) is a manual process by one author and cross-checked by the other. We tried to compare with research papers and the SonarQube set of rules to be as unbiased and accurate as possible, but it is possible that we have misinterpreted or wrongly classified some of the rules.

Lastly, some of the developers' handles were ambiguous, that is, a particular handle could refer to different people as described in Section~\ref{sec:data_analysis}. Therefore, we had to exclude the records of those developers. We also had to ignore the commits and violations that were not tagged as reporter, assignee, or author; and those with developers such as \texttt{GitHub} or \texttt{MergeBot}.

\section{Summary and Future Work}
\label{sec:conclusion}

Overall, the TD dataset offers a rich data source for profiling developers, including
\begin{enumerate}
    \item raw commits to \texttt{main} branches, tagged by author and committer; which allows us to identify and distinguish OSS contributors by the level of contribution
    \item SonarQube diagnostics and closing commits; which allow us to identify the developers and the kinds of diagnostics they contribute and discharge
    \item categorization of refactorings, allowing us to examine the range of refactoring techniques that authors employed: some used very few; others used many.
\end{enumerate}

By analyzing these data sources, we characterize developers by author and commit activity, recognizing those with unusual commit counts compared to authoring. The SonarQube data allows us to profile each developer, distinguishing them by code smells and bad practices. The profiling exercise uncovers evidence supporting maturity and experience. 

Last, the Refactoring Miner data supports that the mature/ex\-per\-ienc\-ed developers employ more sophisticated refactorings, involving generalization and in-depth code knowledge, rather than the simpler renamings that the other developers usually perform.

In profiling developers, we uncover details informing decisions
\begin{itemize}
    \item for assigning pair-programming teams: by knowing the stren\-gths of various developers, teams can be purpose-built to transfer knowledge, or expose new techniques, or to make speedy progress on paramount tasks
    \item for inviting individuals to review-panels to ensure diversity of viewpoints, strengths, and techniques
\end{itemize}

\noindent\textbf{Future Work\ \ \ }
The TD dataset contains additional data sources, including the full details of each commit in \texttt{GIT{\textunderscore}COMMITS{\textunderscore}CHANGES}, and fault-tracking information in \texttt{SZZ{\textunderscore}FAULT{\textunderscore}INDUCING{\textunderscore}COMMITS} tables.  The former would give precise commit data, telling us, on a file-by-file basis, the number of lines inserted, deleted, and moved.  This could sharpen our profiling focus to include expertise in specific packages or code sections.  The latter can be used to identify developers who introduce faults, and the developers who repair them---additional dimensions to profile. The SonarQube data could also be analyzed in another dimension: as a time-series.  Analyzing in this way can allow us to see how developers change their behaviors over time.  Results could inform questions like "do developers become more tolerant to more or different debt as they become more experienced?" and "do developers change the set of repair actions as they become more involved in a project?".  These exciting questions remain for later work. Our data comes from the TD dataset; this decimates the information down to the \texttt{main} branch. In the future, we can draw on additional data sources, including the entire commit log to date, including branches that were prototypes or explored alternate constructions.  Eventually, these profiling activities might form the basis of an actual artifact, a \textit{developer profile}, containing salient details about an individual.

\bibliographystyle{ACM-Reference-Format}
\bibliography{main}

\end{document}